\documentclass[12pt]{article}
\usepackage{amssymb}

\begin{document}

\title{Dibaryons as canonically quantized biskyrmions}
\author{T. Krupovnickas$^1$ E. Norvai\v{s}as$^1$ 
and D.O. Riska$^{2,3}$}
\date{}
\maketitle

\centerline{\it $^1$Institute of Theoretical Physics and Astronomy,
Vilnius, 2600 Lithuania}

\centerline{\it $^2$Helsinki Institute of Physics,
00014 University of Helsinki, Finland}

\centerline{\it $^3$Department of Physics, 00014 University of Helsinki,
Finland}

\vspace{1cm}

\begin{abstract}
The characteristic feature of the ground state configuration of
the Skyrme model description of
nuclei is the 
absence of recognizable individual nucleons. The ground state
of the skyrmion with baryon number 2 is axially symmetric, and is 
well approximated by a simple rational map, which represents a
direct generalization of Skyrme's
hedgehog ansatz for the nucleon.  
If the Lagrangian density is canonically quantized this
configuration may support 
excitations that
lie close and possibly below the threshold for pion decay, and
therefore describe dibaryons. 
The quantum corrections stabilize these solutions, the
mass density of which
have the correct
exponential fall off at large distances.

\end{abstract}
\vspace{1cm}

\newpage
{\bf 1. Introduction}
\vspace{0.5cm}

The ground state solutions to Skyrme's topological soliton model
for baryons \cite{Skyrme61} with baryon numbers that are
larger than 1 have intriguing geometric structures with 
striking polyhedral symmetry
\cite{Battye}. The simplest example is the system with baryon
number 2, which has axial
symmetry \cite{Braaten88}. The deuteron may be viewed as a
quantized version of this ground state 
configuration \cite{Schroers}. It remains an open issue whether
direct semiclassical quantization of these ground state
configurations represent physical systems
\cite{Braaten88,Kopeliovich}. 

The numerical construction 
of these multiskyrmion configurations is a demanding task
\cite{Battye,Walet}. Fortunately is is possible in many cases to
find simple rational
maps, which provide remarkably accurate approximations for the
multiskyrmion ground state configurations \cite{Houghton}. Such
rational maps may be viewed to represent direct formal 
generalizations of Skyrme's
original hedgehog ansatz for the the system with baryon
number 1. 
Employment of the rational map approximation greatly
simplifies the study of the quantized modes of the
multiskyrmion systems \cite{Basker}.
In the case of $B=2$ the ground state solution is
6-dimensional, whereas the physical solution in the
deuteron channel corresponds to a mode in a 12-dimensional
space. We here consider the possibility for dibaryon
solutions other than the deuteron, which may be
represented as spin-isospin excitations of the
ground state solution.

The approach adopted here is to canonically quantize the 
rational map ansatz for the
ground state solution of the baryon number 2 system in
representations of arbitrary dimension for $SU(2)$,
using the method of refs.\cite{Fujii1,Fujii2,Acus97}.
The corresponding states have $I=J$, and therefore
represent dibaryon states other than the deuteron. 
By the conventional semiclassical quantization method 
the lowest one of these states is very deeply bound,
with vibrational excitations that for $J=0$ may lie below
the threshold for decay to two nucleons and a
pion \cite{Basker}.
When canonical {\it ab initio} quantization is employed
the ground state moves to or above the threshold for two-nucleon
decay for all representations. 
In this case the representations of lowest order 
admit several states with $I=J=0,1$, which fall below the 
threshold for pion decay.
From the phenomenological perspective the Skyrme model
thus can accomodate narrow dibaryon states, although 
it does not demand their existence. The overwhelming
experimental evidence is against the existence of
narrow dibaryon states, although some intriguing signals
have been recently been seen in the missing mass
spectrum of the reaction $pd\rightarrow pX$ reaction
\cite{Filkov99,Filkov00}.

This paper is organized into 5 sections. In section 2 the
rational map ansatz for the baryon number 2 skyrmion is
generalized to representations of arbitrary dimension.
This ``biskyrmion'' is canonically quantized in
section 3. In section 4 the resulting equations of
motion are solved numerically, and the spectrum is
obtained. Section 5 contains a concluding discussion.

\vspace{1cm}

{\bf 2. The axially symmetric soliton with $B=2$}

\vspace{0.5cm}

The Skyrme model is a Lagrangian density for a unitary
field $U({\bf r}, t)$ that is described by any
representation of $SU(2)$. 
In an irreducible representation this unitary field
may be   
expressed in terms of three unconstrained Euler angles 
$\vec \alpha=(\alpha ^1,\alpha ^2,\alpha ^3)$ as

\begin{equation}
U({\bf x},t)=D^j(\vec\alpha ({\bf x},t)).
\label{a1}
\end{equation}
Here the elements of the matrices $D^j$ are Wigner D-functions. In an
arbitrary 
reducible representation unitary field can be decomposed into direct sum 
of $D^j$ functions. The Euler angles $\vec \alpha$ then are the dynamical 
variables of the theory.

The Skyrme model is defined by the chirally symmetric
Lagrangian density :
 
\begin{equation}
{\cal L[}U({\bf x},t){\cal ]}=-{\frac{f_\pi ^2}4}{\rm Tr}\{R_\mu 
R^\mu \}+{
\frac 1{32e^2}}{\rm Tr}\{[R_\mu ,R_\nu ]^2\},
\label{a2}
\end{equation}
where the ''right'' current  is defined as
 
\begin{equation}
R_\mu =(\partial _\mu U)U^{\dagger },
\label{a3}
\end{equation}
and $f_\pi $ (the pion decay constant) and $e$ are parameters. 

The rational map ansatz for the Skyrme field with $B=2$ 
(biskyrmion) \cite{Houghton} in the fundamental representation  
may be generalized to an
arbitrary irreducible representation of $SU(2)$ in the following
way:

\begin{equation}
e^{i\left(\hat n\cdot \vec\tau\right)F_R(r)}\Longrightarrow 
U_{R}({\bf r})={\rm{exp}}\{2i\hat n^{a}\cdot \hat{J}_{a}F_R(r)\},  
\label{a4}
\end{equation}
Here $F_R(r)$ is a scalar function (``the chiral angle'') and 
$\hat{\bf n}$ is a unit vector, which may be defined by its circular 
components:

\begin{eqnarray}
&\hat{n}_{+1}=-\hat{n}^{-1}=-\frac{\sin^2\vartheta}{\sqrt{2}
\left(1+\cos^2\vartheta\right)}e^{2\rm{i}\varphi},\nonumber\\
&\hat{n}_{0}=\hat{n}^{0}=\frac{2\cos\vartheta}{1+\cos^2\vartheta},
\nonumber\\
&\hat{n}_{-1}=-\hat{n}^{+1}=\frac{\sin^2\vartheta}{\sqrt{2}
\left(1+\cos^2\vartheta\right)}e^{-2\rm{i}\varphi}.\label{nkomponentes}
\end{eqnarray}

The boundary condition of the chiral angle $F_R(r)$ at the
orgin is $F_R(0)=\pi$.
In terms of Euler angles 
the generalized rational map ansatz may be expressed 
as

\begin{eqnarray}
&\alpha^1=2\varphi-\arctan \left(\frac{2\cos\vartheta\tan 
F_R(r)}{1+\cos^2\vartheta}\right)-\frac\pi2,\nonumber\\
&\alpha^2=-2\arcsin 
\left(\frac{\sin^2\vartheta\sin 
F_R(r)}{1+\cos^2\vartheta}\right),\nonumber\\
&\alpha^3=-2\varphi-\arctan 
\left(\frac{2\cos\vartheta\tan 
F_R(r)}{1+\cos^2\vartheta}\right)+\frac\pi2.\label{toreuler}
\end{eqnarray}

Given the rational map ansatz (\ref{a4}), the Lagrangian density (\ref{a2}) 
reduces to the following rather simple form

\begin{eqnarray}
&&{\cal L}=-\frac13j(j+1)(2j+1)\Biggl\{f_\pi^2\left[F_R^{\prime 2}(r)+
8\frac{\sin^2\vartheta\sin^2F_R(r)}{r^2(1+
\cos^2\vartheta)^2}\right]\nonumber\\
&&+\frac{8}{e^2}\frac{\sin^2\vartheta\sin^2F_R(r)}{r^2(1+
\cos^2\vartheta)^2}\left[F_R^{\prime 
2}(r)+2\frac{\sin^2\vartheta\sin^2F_R(r)}{r^2(1+
\cos^2\vartheta)^2}\right]\Biggr\}.
\label{klastorlag}
\end{eqnarray}
The classical Lagrangian density depends on the 
dimension of the representation $j$ only through the overall scalar factor
$N={2\over 3}j(j+1)(2j+1)$, which can be absorbed by a renormalization
of the parameters \cite{Norvaisas}.

In contrast to hedgehog form for the $B=1$ skyrmion
the rational map Lagrangian for the skyrmion with $B=2$
depends on both the polar angle 
$\vartheta$ and the radius $r$. Integration and normalization 
yields the following expression for the mass of the biskyrmion:

\begin{eqnarray}
&&M_0(F_R)=2\pi\frac{f_\pi}{e}\int^{\infty}_{0}d\tilde 
r\Biggl\{\tilde r^2F_R^{\prime 2}(\tilde r)
\left(1+4\frac{\sin^2F_R(\tilde r)}{\tilde r^2}\right)\nonumber\\
&&+\sin^2F_R(\tilde r)\left(4+\left(\frac83+\pi\right)
\frac{\sin^2F_R(\tilde r)}{\tilde r^2}\right)\Biggr\}
\label{klastormase}.
\end{eqnarray}
Here the dimensionless variable $\tilde r$ is defined as 
$\tilde r=ef_\pi r$. Variation of this 
expression for the mass leads to the differential 
equation for the ``chiral angle'':

\begin{eqnarray}
&&F_R^{\prime\prime}(\tilde r)\left(1+4\frac{\sin^2{F_R(\tilde r)}}{\tilde r^2}
\right)+2F_R^{\prime 2}(\tilde r)
\frac{\sin{2F_R(\tilde r)}}{\tilde r^2}+\frac{2}{\tilde r}
F_R^\prime(\tilde r)\nonumber\\
&&-\frac{\sin2F_R(\tilde r)}{\tilde r^2}
\left(2+\left(\frac83+\pi\right)\frac{\sin^2F_R(\tilde r)}{\tilde r^2}
\right)=0.
\label{klastorlygtis}
\end{eqnarray}
At large distances $\tilde r\rightarrow\infty$ 
this equation has the asymptotic form:

\begin{equation}
F_R^{\prime\prime}(\tilde r)+\frac{2}{\tilde r}F_R^\prime(\tilde r)
-\frac{4}{\tilde r^2}F_R(\tilde r)=0.
\label{klastoraslygt}
\end{equation}
This equation differs from the corresponding asymptotic equation
for the chiral angle of the B=1 hedhehog solution only in the
value of the coefficient in the numerator of the last term
on the l.h.s. (4 instead of 2). 
The solution of this asymptotic equation (\ref{klastoraslygt}) is

\begin{equation}
F_R(\tilde r)=C\tilde r^{-\frac{1+\sqrt{17}}{2}}.
\label{asimptspr}
\end{equation}
The fall off rate is somewhat larger here than in the case of
B=1, as the power of $\tilde r$ in (12) is -2.56 whereas it
in the case B=1 is -2. 
\vspace{1cm}

{\bf 3. Quantization of the biskyrmion}
\vspace{0.5cm}

We shall employ collective rotational coordinates to separate the variables 
which depend on the time and spatial coordinates:

\begin{equation}
U({\bf r},{\bf q}(t))=A\left({\bf q}(t)\right) U_0({\bf r}) 
A^\dagger \left({\bf q}(t))\right).
\label{b1}
\end{equation}
We consider the Skyrme model quantum mechanically {\it ab initio} and 
thus treat the 
generalized (collective) coordinates $\bf q(t)$ and the
corresponding velocities 
$\dot{\bf q}$ as dynamical variables, which satisfy the 
commutation relations

\begin{equation}
\bigl[\dot 
q^\alpha,q^\beta\bigr]=-{\rm{i}}\,{{}_R f^{\alpha\beta}({\bf q})}.
\label{b2}
\end{equation}
The functions $_R f^{\alpha\beta}$ depend on the generalized
coordinates {\bf q}, the explicit form of which is determined by the canonical
commutation relations below \cite{Acus98}.
After substitution 
of (\ref{b1}) into the Lagrangian density (\ref{a2}) the 
dependence of Lagrangian 
on the generalized velocities may be expressed as

\begin{eqnarray}
\hat{L}(\dot {\bf q},{\bf q},F_R)=&\frac1N \int \ 
\hat{\!\!\!\cal L}\bigl({\bf n},\dot{\bf q} (t),{\bf q}(t),F_R(r)\bigr)
r^2\sin\vartheta dr d\vartheta d\varphi \nonumber \\
=& \frac 12\dot q^\alpha {}_Rg_{\alpha\beta}({\bf q})\dot q^\beta
+{\cal O}(\dot q^0 ).
\label{b3}
\end{eqnarray}
Here we have used the notation
\begin{eqnarray}
{}_Rg_{\alpha\beta}({\bf q})&=&\sum_m(-)^m
C_\alpha^{\prime
(m)}({\bf q}){}_Ra_m(F_R)C_\beta^{\prime(-m)}({\bf q})\nonumber\\
 &=&\left(\begin{array}{ccc}
2(a_0\cos^2q^2+a_1\sin^2q^2) & 0&2a_0\cos q^2\\0 & 2a_1&0\\2a_0\cos
q^2&0&2a_0\end{array}\right) .
\label{b4}
\end{eqnarray}
The functions of dynamical variables $C_\alpha^{\prime (m)}({\bf q})$ are 
defined in \cite{Acus97}. 

Because of the axial symmetry of the rational map configuration
there are only 
two different moments of inertia, which may be defined as:

\begin{eqnarray}
&&a_0={}_Ra_0(F_R)=\nonumber\\
&&\frac{\pi}{3e^3f_\pi}\int^{\infty}_{0}d\tilde 
r\tilde r^2\sin^2F_R\left((12-3\pi)(1+F_R^{\prime 
2})+8\frac{\sin^2F_R}{\tilde r^2}\right),
\label{b5}\\
&&a_1={}_Ra_1(F_R)={}_Ra_{-1}(F_R)\nonumber\\
&&=\frac{\pi}{3\sqrt{2}e^3f_\pi}\int^{\infty}_{0} d\tilde 
r\tilde r^2\sin^2F_R\left(3\pi(1+F_R^{\prime 
2})+16\frac{\sin^2F_R}{\tilde r^2}\right).
\label{b6}
\end{eqnarray}

The generalized momentum operators, which are canonically
conjugate to ${\bf q}$, are defined as

\begin{equation}
p_\alpha=\frac{\partial L}{\partial\dot q^\alpha}=\frac12\bigl\{\dot 
q^\beta,{}_Rg_{\alpha\beta}({\bf q})\bigr\}.
\label{b7}
\end{equation}
These operators satisfy the canonical commutation relations

\begin{equation}
\bigl[p_\alpha,q^\beta\bigr]=-\rm{i}\, \delta_{\alpha\beta},
\label{b8}
\end{equation}
from which it follows that the explicit form for the matrix 
${}_Rf^{\alpha\beta}({\bf q},F_R)$ in eq. (\ref{b2})
is

\begin{equation}
{}_Rf^{\alpha\beta}({\bf q},F_R)=C_{(a)}^{\prime\alpha}({\bf q})
\left(a^{-1}(F_R)
\right)^{ab}C_{(b)}^{\prime\beta}({\bf q}).
\label{b9}
\end{equation}

The group parameter manifold of $SU(3)$ is the 
hypersphere $S^3$. It is convenient to 
introduce the 
following angular momentum operators \cite{Acus97} on the sphere:

\begin{eqnarray}
\hat{J}^{\prime b}=
\frac{\rm{i}}{\sqrt{2}}\left\{ p_{\beta},C_{\left(b\right)}^{\prime \beta}
({\bf q})\right\} =(-)^{b}\frac{\rm{i}}{\sqrt{2}}a_b(F_R)\left\{ \dot{q}
^{\beta},C_{\beta}^{\prime \left( -b\right) }({\bf q} \right\} ,
\label{b9a}
\end{eqnarray}
The components of this operator 
satisfy the standard commutation relations. Note that in
eq. (\ref{b9a}) there is no summation over the index $b$.

By some lengthy manipulation the Lagrangian (\ref{a2}) brought into the 
following explicit form

\begin{eqnarray}
&\hat{L}\left(\dot{\bf q},{\bf q},F_R\right)=-M_0(F_R)-\Delta 
M_{j}(F_R)+\frac14\left[\frac{1}{a_1}\hat{J}^{\prime 2}
+\left(\frac{1}{a_0}-
\frac{1}{a_1}\right)\hat{J}^{\prime 2}_0\right].\label{b10}
\end{eqnarray}
Here $\Delta M_{j}(F_R)$ 
represents the quantum mass correction, which may be written as

\begin{eqnarray}
&
\Delta M_{j}(F_R)
=\frac{1}{a_0^2}\Delta M_{0}+\frac{1}{a_0a_1}\Delta 
M_{01}+\frac{1}{a_1^2}\Delta M_{1}.
\label{b11}
\end{eqnarray}
Here the three terms on the r.h.s are quantum corrections that are due 
to the different moments of inertia, which may be written in the
form

\begin{eqnarray}
&\Delta M_{0}=\frac{1}{e^3f_\pi}\Delta\tilde M_{0},\quad\quad
\Delta M_{01}=\frac{1}{e^3f_\pi}\Delta\tilde M_{01},\quad\quad
\Delta M_{1}=\frac{1}{e^3f_\pi}\Delta\tilde M_{1}.
\end{eqnarray}
The dimensionless moments of inertia here are defined as

\begin{eqnarray}
&\Delta\tilde M_{0}=-\frac{\pi}{\sqrt{2}}\int^{\infty}_{0} d\tilde r 
\tilde 
r^2\Biggl\{\frac{4-\pi}{8}\sin^2F_R+\frac{32-9\pi}{80}\left(2j-1\right)
\left(2j+3\right)\sin^4F_R\nonumber\\
&+\frac{F_R^{\prime 
2}\sin^2F_R}{80}\Bigl[8\left(16j\left(j+1\right)-7\right)-\pi
\left(36j\left(j+1\right)-17\right)\nonumber\\
&-\frac{2}{3}\left(32-9\pi\right)
\left(2j-1\right)\left(2j+3\right)\sin^2F_R\Bigr]
+\frac{32j\left(j+1\right)+1}{75}\frac{\sin^4F_R}{\tilde 
r^2}\Biggr\}\\
&\Delta\tilde M_{01}=-\sqrt{2}\pi\int^{\infty}_{0} d\tilde r 
\tilde 
r^2\Biggl\{\frac{4-\pi}{8}\sin^2F_R+\frac{\pi}{80}\left(2j-1\right)
\left(2j+3\right)\sin^4F_R\nonumber\\
&+\frac{F_R^{\prime 
2}\sin^2F_R}{80}\Bigl[40+\pi
\left(4j\left(j+1\right)-13\right)\nonumber\\
&-\frac{2}{3}\left(16-3\pi\right)\left(2j-1\right)\left(2j+3\right)\sin^2F_R
\Bigr]-\frac{\left(2j-1\right)\left(2j+3\right)}{75}\frac{\sin^4F_R}
{\tilde r^2}\Biggr\}\\
&\Delta\tilde M_{1}=-\frac{\pi}{\sqrt{2}}\int^{\infty}_{0} d\tilde r 
\tilde 
r^2\Biggl\{\frac{3\pi-4}{8}\sin^2F_R+\frac{7\pi}{80}\left(2j-1\right)
\left(2j+3\right)\sin^4F_R\nonumber\\
&+\frac{F_R^{\prime 
2}\sin^2F_R}{80}\Bigl[-40+\pi
\left(28j\left(j+1\right)+9\right)\nonumber\\
&-\frac{2}{3}\left(15\pi-16\right)
\left(2j-1\right)\left(2j+3\right)\sin^2F_R\Bigr]
+\frac{52j\left(j+1\right)+11}{75}\frac{\sin^4F_R}{\tilde 
r^2}\Biggr\}.
\end{eqnarray}

The normalized eigenstates with 
fixed spin and isospin $\ell$ of
the corresponding Hamilton operator are

\begin{equation}
\left|
\begin{array}{c}
\ell  \\
{m,m^{\prime }}
\end{array}
\right\rangle =\frac{\sqrt{2\ell +1}}{4\pi }D_{m,m^{\prime }}^{\ell }
({\bf q})\left| 0\right\rangle . 
\label{b12}
\end{equation}
The eigenvalues that correspond to these states give the masses of the
quantum biskyrmion states as:
\begin{eqnarray}
&M_R=M_0(F_R)+\Delta M_{j}(F_R)
+\frac14\left[\frac{1}{a_1}\ell(\ell+1)
+\left(\frac{1}{a_0}-\frac{1}{a_1}\right)m^2_t\right].
\label{b13}
\end{eqnarray}
Here $m_t$ is the third component of the isospin.
The chiral angle $F_R$ is determined by the solution of the
integrodifferential equation that is obtained by
minimization of this expression (\ref{b13}) for the biskyrmion
mass:
\newpage
\begin{eqnarray}
&F_R^{\prime\prime}\Biggl\{-2\tilde r^2-8\sin^2F_R+e^4\tilde 
r^2\sin^2F_R\Bigl[-\frac{\pi-4}{\tilde a_0^3\tilde a_1}\left(2\tilde a_1
\Delta\tilde M_{0}+\tilde a_0\Delta\tilde M_{01}\right)\nonumber\\
&+\frac{\pi}{2\tilde a_0\tilde a_1^3}\left(\tilde a_1
\Delta\tilde M_{01}+2\tilde a_0\Delta\tilde M_{1}\right)+\frac{8\left(
16j\left(j+1\right)-7\right)-\pi\left(36j\left(j+1\right)-17\right)}
{160\tilde a_0^2}\nonumber\\
&+\frac{40+\pi\left(4j\left(j+1\right)-13\right)}
{80\tilde 
a_0\tilde 
a_1}+\frac{-40+\pi\left(28j\left(j+1\right)+9\right)}{160\tilde 
a_1^2}+\frac\pi8\frac{\ell\left(\ell+1\right)-m^2}{\tilde 
a_1^2}\nonumber\\
&+\left(1-\frac\pi4\right)\frac{m^2}{\tilde 
a_0^2}+\frac{\left(2j-1\right)
\left(2j+3\right)}{240}\sin^2F_R\left(\frac{9\pi-32}{\tilde 
a_0^2}+\frac{2\left(3\pi-16\right)}{\tilde a_0\tilde 
a_1}+\frac{16-15\pi}{\tilde a_1^2}\right)\Bigr]\Biggr\}\nonumber\\
&+F_R^{\prime 2}\Biggl\{-4\sin 2F_R+e^4\tilde 
r^2\sin 2F_R\Bigl[-\frac{\pi-4}{2\tilde a_0^3\tilde a_1}\left(2\tilde a_1
\Delta\tilde M_{0}+\tilde a_0\Delta\tilde M_{01}\right)\nonumber\\
&+\frac{\pi}{4\tilde a_0\tilde a_1^3}\left(\tilde a_1
\Delta\tilde M_{01}+2\tilde a_0\Delta\tilde M_{1}\right)+\frac{8\left(
16j\left(j+1\right)-7\right)-\pi\left(36j\left(j+1\right)-17\right)}
{320\tilde a_0^2}\nonumber\\
&+\frac{40+\pi\left(4j\left(j+1\right)-13\right)}
{160\tilde 
a_0\tilde 
a_1}+\frac{-40+\pi\left(28j\left(j+1\right)+9\right)}{320\tilde 
a_1^2}+\frac{\pi}{16}\frac{\ell\left(\ell+1\right)-m^2}{\tilde 
a_1^2}\nonumber\\
&+\frac12\left(1-\frac\pi4\right)\frac{m^2}{\tilde 
a_0^2}+\frac{\left(2j-1\right)
\left(2j+3\right)}{240}\sin^2F_R\Bigl(\frac{9\pi-32}
{\tilde a_0^2}+\frac{2\left(3\pi-16\right)}{\tilde 
a_0\tilde a_1}\nonumber\\
&+\frac{16-15\pi}{\tilde 
a_1^2}\Bigr)\Bigr]\Biggr\}+F_R^{\prime}\Biggl\{-4\tilde r+e^4\tilde 
r\sin^2F_R\Bigl[\frac{-2\left(\pi-4\right)}{\tilde a_0^3\tilde 
a_1}\left(2\tilde a_1\Delta\tilde M_{0}+\tilde a_0\Delta\tilde 
M_{01}\right)\nonumber\\&+\frac{\pi}{\tilde a_0\tilde a_1^3}\left(\tilde a_1
\Delta\tilde M_{01}+2\tilde a_0\Delta\tilde M_{1}\right)+\frac{8\left(
16j\left(j+1\right)-7\right)-\pi\left(36j\left(j+1\right)-17\right)}
{80\tilde a_0^2}\nonumber\\
&+\frac{40+\pi\left(4j\left(j+1\right)-13\right)}
{40\tilde 
a_0\tilde 
a_1}+\frac{-40+\pi\left(28j\left(j+1\right)+9\right)}{80\tilde 
a_1^2}+\frac\pi4\frac{\ell\left(\ell+1\right)-m^2}{\tilde 
a_1^2}\nonumber\\
&+2\left(1-\frac\pi4\right)\frac{m^2}{\tilde 
a_0^2}+\frac{\left(2j-1\right)
\left(2j+3\right)}{120}\sin^2F_R\Bigl(\frac{9\pi-32}
{\tilde a_0^2}+\frac{2\left(3\pi-16\right)}{\tilde 
a_0\tilde a_1}\nonumber\\
&+\frac{16-15\pi}{\tilde a_1^2}\Bigr)\Bigr]\Biggr\}+\sin 
2F_R\Biggl\{4+\frac23\left(3\pi+8\right)\frac{\sin^2F_R}{\tilde 
r^2}+e^4\Bigl[-\frac{\left(12-3\pi\right)\tilde r^2+16\sin^2F_R}{6\tilde 
a_0^3\tilde a_1}\nonumber\\
&\times\left(2\tilde a_1\Delta\tilde M_{0}+\tilde 
a_0\Delta\tilde M_{01}\right)-\frac{3\pi\tilde 
r^2+32\sin^2F_R}{12\tilde a_0\tilde a_1^3}\left(\tilde a_1\Delta\tilde 
M_{01}+2\tilde a_0\Delta\tilde M_{1}\right)\nonumber\\
&+\frac{\tilde 
r^2}{32}\left(\frac{\pi-4}{\tilde 
a_0^2}+\frac{2\left(\pi-4\right)}{\tilde 
a_0\tilde 
a_1}+\frac{4-3\pi}{\tilde 
a_1^2}-2\pi\frac{\ell\left(\ell+1\right)-m^2}{\tilde 
a_1^2}-16\left(1-\frac\pi4\right)\frac{m^2}{\tilde 
a_0^2}\right)\nonumber\\
&-\frac{\sin^2F_R}{150}\Bigl(\frac{32j\left(j+1\right)+1}{\tilde 
a_0^2}-\frac{2\left(2j-1\right)\left(2j+3\right)}{\tilde 
a_0\tilde 
a_1}+\frac{52j\left(j+1\right)+11}{\tilde 
a_1^2}\nonumber\\
&+100\frac{\ell\left(\ell+1\right)-m^2}{\tilde 
a_1^2}+100\frac{m^2}{\tilde 
a_0^2}\Bigr)\nonumber\\
&-\frac{\left(2j-1\right)\left(2j+3\right)}{160}\tilde 
r^2\sin^2F_R\left(\frac{32-9\pi}{\tilde a_0^2}+\frac{2\pi}{\tilde 
a_0\tilde 
a_1}+\frac{7\pi}{\tilde 
a_1^2}\right)\Bigr]\Biggr\}=0.
\label{b14}
\end{eqnarray}
Here the mass parameters $\Delta M_j$ and the moments
of inertia are integrals of the chiral angle that is
determined by solution.

At large distances the equation (\ref{b14}) reduces to the asymptotic form

\begin{eqnarray}
&\tilde r^2F_R^{\prime\prime}+2\tilde 
rF_R^\prime-\left(4+\tilde m^2\tilde r^2\right)F_R=0.
\label{b15}
\end{eqnarray}
Here the quantity $\tilde m$ is defined as

\begin{eqnarray}
\tilde m^2=e^4\Bigl[&-\frac{12-3\pi}{6\tilde a_0^3\tilde 
a_1}\left(2\tilde a_1\Delta\tilde M_{0}+\tilde 
a_0\Delta\tilde M_{01}\right)-\frac{3\pi}{12\tilde 
a_0\tilde a_1^3}\left(\tilde a_1\Delta\tilde M_{01}+2\tilde 
a_0\Delta\tilde M_{1}\right)\nonumber\\
&-\frac{1}{32}\Bigl(\frac{4-\pi}{\tilde 
a_0^2}+\frac{2\left(4-\pi\right)}{\tilde 
a_0\tilde 
a_1}+\frac{3\pi-4}{\tilde 
a_1^2}+2\pi\frac{\ell\left(\ell+1\right)-m^2}{\tilde 
a_1^2}\nonumber\\
&+16\left(1-\frac\pi4\right)\frac{m^2}{\tilde 
a_0^2}\Bigr)\Bigr].
\label{b16}
\end{eqnarray}
This mass parameter describes the behavior of the chiral angle at infinity:

\begin{eqnarray}
&F_R\left(\tilde 
r\right)=C\left(\frac{\tilde m}{\tilde r}+\frac{2}{\tilde r^2}\right)
e^{-\tilde m\tilde r}.
\label{b17}
\end{eqnarray}
The quantity $m=ef_{\pi}\tilde m$ represents an effective pion
mass, which governs the asymptotic fall off
$\exp (-2mr)$ of biskyrmion mass density. In the case
of the $B=1$ skyrmion the corresponding asymptotic
fall off is $\exp(-mr)$ and represents the 
Yukawa form of the pion cloud around the nucleon.

\vspace{1cm}

{\bf 4. Numerical results}
\vspace{0.5cm}

We have solved numerically the integrodifferential 
equation (\ref{b14})  for the
chiral angle of the rational map ansatz for the biskyrmion
(\ref{a4}), which provides a good approximation to the ground state
solution for the $B=2$ skyrmion. 
The resulting biskyrmion mass values are given in Table 1. 
In the numerical calculation we employed the 
same values for the two parameters
of the the model, $f_\pi$ and $e$, as were obtained in 
ref.\cite{Acus98} by fitting the empirical values for mass
and the the isoscalar radius of the nucleon. We also solved the eq. 
(\ref{klastorlygtis}) for the classical case using the same
parameter values. The calculated values for the nucleon
mass in the classical case are $978$ MeV, $1028$ MeV
and $1090$ MeV respectively for the three sets of parameter values
used below to reproduce the empirical value $939$ MeV 
with canonical quantization.

In the classical treatment the rational map ansatz leads to 
a deeply bound 
biskyrmion
solution with $I=J=0$ that is stable against decay to two nucleons. This
is indicated the negativity of the parameter $\Delta M = M-2M_N$,
where $M$ is the biskyrmion mass and $M_N$ is the calculated 
mass of the nucleon. In the canonically quantized case this
state moves up to or above the threshold for nucleon
decay. In the fundamental representation the $I=J=0$ state
is marginally
stable against decay to two nucleons with the present
choice of parameter values.
This is not the deuteron state, which has spin 1 and isospin 0. In view
of the very small energy by which this solution falls below the
sum of two nucleon masses (- 5 MeV) and 
the absence of any such state in all other
representations of $SU(2)$ we believe interpret this result to
be an accidental consequence of the approximate character of the
rational map ansatz, and thus that, within the margin of error,
there is no bound state. 

The $\ell =I=J=0$ state is found to lie below the threshold for
decay into two nucleons and a pion in the canonically quantized case
in both the fundamental representation and the three dimensional
representation ($j=1$). 
There are three states with $\ell =I=J=1$ in the $j=1$ representation, 
which lie below the 
threshold of pion decay of the dibaryon. 
With the rational map ansatz the energies of the state
with $I=J=0$ is found to be roughly 1950~MeV and
the energies of the states with $I=J=1$ and isospin projection
$m_t=0$ at 2000~MeV and with $m_t=\pm1$ at 2010~MeV respectively.
In the 4 dimensional representation $j=3/2$ all these states
are found to lie well above the threshold for pion decay,
when the same parameter values are used. Should the 
recent empirical indications \cite{Filkov00} for three 
"supernarrow" dibaryons at 1904~MeV, 1926~MeV and 1942~MeV be
confirmed, they could thus be accomodated within the Skyrme
model framework, as long as their spin and isospin equal 0,
1, 1 respectively..
The $\gamma$ decay pattern of those states, do however
suggest
that they all have isospin 1 and $J^P=1^\pm$ 
\cite{Filkov99,Tatis}.

Typical contours of constant 
classical and quantum mass density of the dibaryon are plotted in Fig.1
for the case $j = 1/2$ and $\ell = 0$. 
The corresponding
typical contours of constant baryon number density are shown in Fig.2.  
Comparison of the classical and quantum solutions 
in Figs.1 and 2 show that the 
quantum soliton is concentrated is more compact 
as the densities fall off exponentially at infinity
in for the latter. 

\begin{table}[h]
\caption{Dibaryon parameter dependence on $SU(2)$ group 
representation\newline}
\label{tablas}
\begin{tabular}{c|c|c|c|c|c|c|c}
\hline \hline
$j$ & $\ell$ & $m_t$ & $e$ & $f_\pi$ (MeV) & $M$ (MeV) & $\Delta M$ (MeV)  & 
$m$ (MeV) \\\hline\hline
$\scriptstyle\frac12$ & Classical & & 4.46 & 59.8 & 1918.5 & -36.9 &\\\hline
$\scriptstyle\frac12$ & 0 & 0 & 4.46 & 59.8 & 1873.0 & -5.0 & 
62.9\\\hline\hline
1 & Classical & & 4.15 & 58.5 & 2017.0 & -38.8
&\\\hline
1 & 0 & 0 & 4.15 & 58.5 & 1949.9 & 71.9 & 88.1\\\hline
1 & 1 & 0 & 4.15 & 58.5 & 1998.5 & 120.5 & 53.7\\\hline
1 & 1 & $\pm1$ & 4.15 & 58.5 & 2012.0 & 134.0 & 45.9\\\hline\hline
$\scriptstyle\frac32$ & Classical & & 3.86 & 57.7 & 2138.9 & -41.1 & 
\\\hline
$\scriptstyle\frac32$ & 0 & 0 & 3.86 & 57.7 & 2049.7 & 171.7 & 
100.6\\\hline
$\scriptstyle\frac32$ & 1 & 0 & 3.86 & 57.7 & 2090.5 & 
212.5 & 78.8
\\\hline$\scriptstyle\frac32$ & 1 & $\pm1$ & 3.86 & 57.7 & 
2101.8 & 223.8 & 74.6
\\\hline\hline
\end{tabular}
\end{table}




\

\vspace{1cm}

{\bf 5. Discussion}
\vspace{0.5cm}

The present work represents an exploratory calculation of the
possible quantum excitations that the the axisymmetric
ground state configuration of the $B=2$ skyrmion may
support, once self consistent canonical quantization is
imposed. We have here considered the possible 
states with $I=J$, which represent dibaryons other than 
the deuteron, which has been considered in ref.
\cite{Schroers}. 

The calculation is based on the rational map ansatz of
ref.\cite{Houghton}, which provides a good approximation
to the actual ground state solution. The calculation, 
within the margin of error that is associated with the
rational map approximation does not yield states that
would be stable against decay to two nucleons. One
dibaryon state with
$I=J=0$ that is stable against decay into two nucleons
and a pion appears in both the fundamental two dimensional
as well as in the the three dimensional representation.
The three dimensional representation also may accomodate
two such dibaryon states with $I=J=1$.
It is of
course intriguing that there are some recent 
empirical indications for such states \cite{Filkov00},
and that for those $I=1$ would be favored by the
photon decay pattern \cite{Tatis}.   

The possibility for a dibaryon state with $I=0$ has drawn
considerable experimental and theoretical interest over the
past 10 years, but no confirmed evidence for such has yet
been found \cite{Bilger}. Should in the end no such 
dibaryon state be found, it would put constraints on 
the choice of representation to be used with the
rational map ansatz in the canonically quantized case, and
require employment of
dimension greater greater than or equal to 4, as in such
the dibaryon states lie above the threshold of pion decay.

\vspace{1cm}
{\bf Acknowledgment} \vspace{0.5cm}

The research of T.K. and E.N. 
was made possible in part by the Grant 
No 99235 from the Lithuanian State Science and Studies Foundation.
Research supported in part by the Academy of Finland by
grants No. 43982 and 44903.

\vspace{2cm}

\vspace{1cm}

{\bf Figure Captions}

\vspace{0.5cm}

Fig.1 Typical contours of constant classical (dashed line) and 
quantum (solid line) mass densities of the
biskyrmion $j=\frac12$, $\ell =0$ case.

\vspace{0.5cm}

Fig.2 Typical contours of constant classical (dashed line) and 
quantum (solid line) baryon number densities of the
biskyrmion $j=\frac12$, 
$\ell =0$ case.

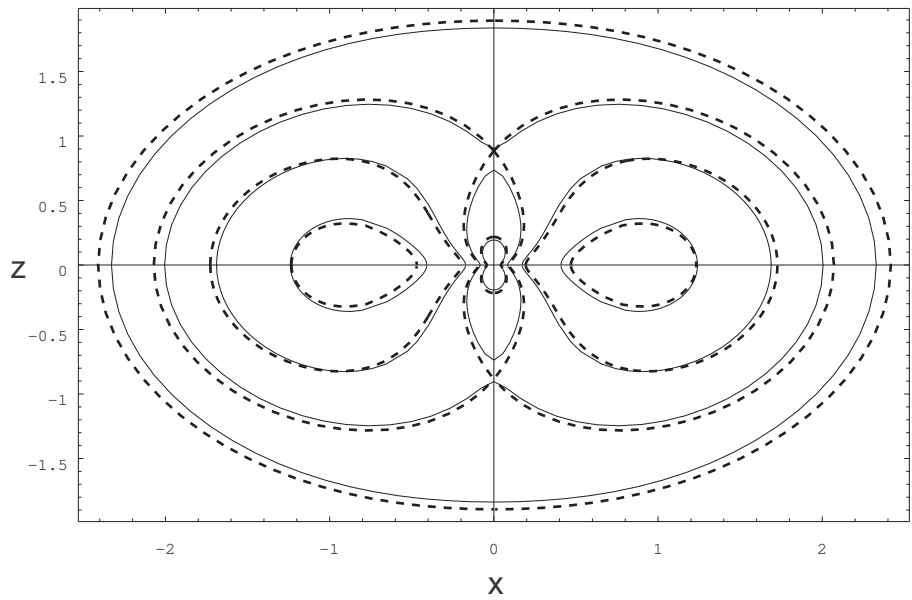
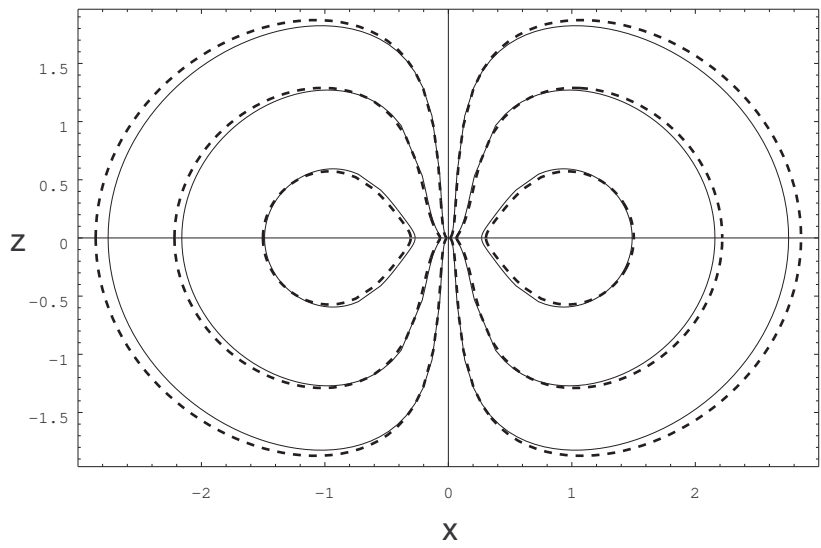
\end{document}